\documentclass[a4paper,pra,reprint,twocolumn,superscriptaddress]{revtex4-2}

\usepackage{ulem}
\usepackage{amssymb}
\usepackage{amsmath}
\usepackage{epsfig}
\usepackage{color}
\usepackage{graphics, graphicx}
\usepackage{bbold}
\usepackage{psfrag}
\usepackage{mathcomp}
\usepackage{amsmath}
\usepackage{amssymb}%������ĸ�Ӵ�
\usepackage{mathrsfs}%������ĸ
\usepackage{subfigure}
\usepackage{verbatim}
\usepackage{xcolor,bm}
\usepackage[colorlinks,citecolor=blue,urlcolor=blue]{hyperref}

\def\bra#1{\left<{#1}\right|}				% "bra"-state
\def\ket#1{\left|{#1}\right>}				% "ket"-state
\usepackage{mathrsfs}

\begin{document}

\title{Characterizing the Yang-Lee zeros of the classical Ising model through dynamic quantum phase transitions}
\author{Mingtao Xu}
\affiliation{University of Science and Technology of China, Hefei 230088, China}
\author{Wei Yi}
\email{wyiz@ustc.edu.cn}
\affiliation{CAS Key Laboratory of Quantum Information, University of Science and Technology of China, Hefei 230026, China}
\affiliation{Anhui Province Key Laboratory of Quantum Network, University of Science and Technology of China, Hefei, 230026, China}
\affiliation{CAS Center For Excellence in Quantum Information and Quantum Physics, Hefei 230026, China}
\affiliation{Hefei National Laboratory, University of Science and Technology of China, Hefei 230088,
China}
\author{De-Huan Cai}
\email{dhcai@ustc.edu.cn}
\affiliation{Hefei National Laboratory, University of Science and Technology of China, Hefei 230088,
China}
\affiliation{Bengbu University, Bengbu 233030, China }

\begin{abstract}
In quantum dynamics, the Loschmidt amplitude is analogous to the partition function in the canonical ensemble. Zeros in the partition function indicate a phase transition, while the presence of zeros in the Loschmidt amplitude indicates a dynamical quantum phase transition. Based on the classical-quantum correspondence, we demonstrate that the partition function of a classical Ising model is equivalent to the Loschmidt amplitude in non-Hermitian dynamics, thereby mapping an Ising model with variable system size to the non-Hermitian dynamics. It follows that the Yang-Lee zeros and the Yang-Lee edge singularity of the classical Ising model correspond to the critical times of the dynamic quantum phase transitions and the exceptional point of the non-Hermitian Hamiltonian, respectively. Our work reveals an inner connection between Yang-Lee zeros and non-Hermitian dynamics, offering a dynamic characterization of the former.

\end{abstract}

\maketitle

\section{Introduction}
\begin{figure}[tbp]
  \centering
  \includegraphics[width=9cm]{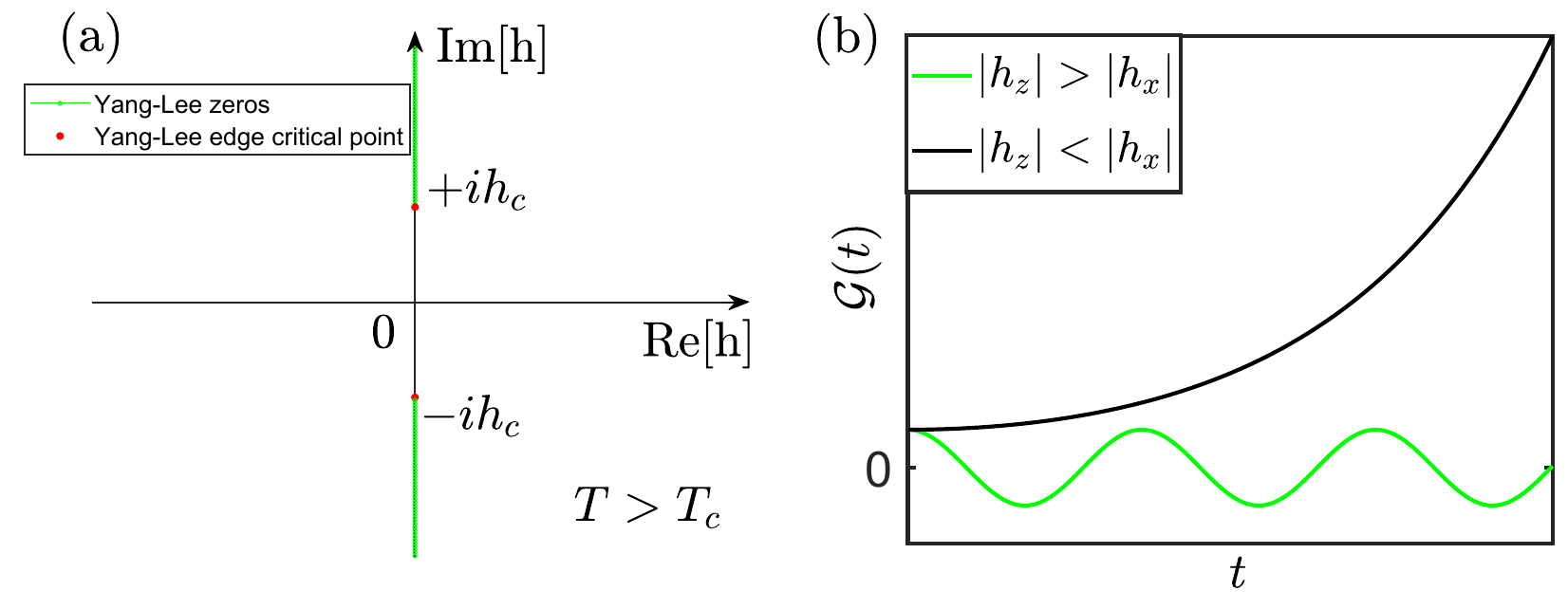}
  \caption{(a) The phase diagram of the classical Ising model; (b) The Loschmits amplitude of non-Hermitian quantum system for two cases: dynamic quantum phase transitions exist in the case of $|h_z|>|h_x|$, while dynamic quantum phase transitions does not exist in the case of $|h_z|<|h_x|$.}
  \label{fig:fig1}
\end{figure}

The Yang-Lee zeros~\cite{Yang,Lee} are the zero points of the partition function in the complex plane of physical parameters. In the thermodynamic limit, the zeros become densely distributed in the complex parameter plane, and it signifies the presence of a phase transition when the zeros cross the real axis~\cite{Yang,Lee}, during which thermodynamic quantities may exhibit non-analytic behavior. In the classical ferromagnetic Ising model, the Yang-Lee zeros distribute on the imaginary axis of the complex plane formed by an external magnetic field~\cite{Simon,Newman,Lieb}, and there exists a nonzero lower bound for the distribution of the zeros in the disordered phase, as shown in Fig.~\ref{fig:fig1}(a). In the vicinity of the lower bound, that is, at the edge of the distribution, a critical phenomenon called the Yang-Lee edge singularity~\cite{Kortman,Fisher1,Kurtze,Fisher2,Cardy1,Cardy2,Zamolodchikov,Kist,Vecsei} arises, which is accompanied by anomalous scaling laws~\cite{Itzykson1,Itzykson2,Couvreur,Chang}.

Although Yang-Lee zeros and Yang-Lee edge singularity have been extensively studied~\cite{Binek1,Binek2,Wei1,Peng,Brandner,Deger1,Deger2,Connelly,Jian,Heyl1,Wei2,Wei3,Li1,Li2,Lu,Timonin}, with experimental observations of Yang-Lee zeros~\cite{Binek1,Binek2,Wei1,Peng,Brandner} already achieved, direct observation of the Yang-Lee edge singularity and its associated anomalous scaling laws remains challenging due to the requirement of an imaginary magnetic fields. It was not until recently that Matsumoto \textit{et al}.~\cite{Matsumoto} proposed that the Yang-Lee edge singularity can be implemented in open quantum systems. Their approach was based on the quantum-classical correspondence, where the partition function of a classical one-dimensional(1D) Ising model with an imaginary magnetic field was shown to be equivalent to the partition function of a non-Hermitian quantum system in the continuum limit. Subsequently, experimental measurements of Yang-Lee zeros and anomalous scaling behavior are reported in a heralded single-photon system through non-unitary evolution~\cite{Gao}, thereby confirming the theoretical predictions regarding the Yang-Lee edge singularity.

Here, we revisit the partition function of the classical 1D Ising model. When the applied magnetic field exceeds the critical value, the system's partition function takes the form of a cosine function and Yang-Lee zeros appear, while below the critical magnetic field, the zeros vanish. This is very similar to the properties of the Loschmidt amplitude in the quantum dynamics driven by a non-Hermitian effective Hamiltonian~\cite{Bender1,Bender2,Bender3,Ganainy,Choi,Fang}. As shown in Fig.~\ref{fig:fig1}(b), when $|h_z|>|h_x|$, the Loschmidt amplitude periodically exhibits zeros, indicating that the system undergoes a dynamic quantum phase transition~\cite{Heyl1,Heyl2,Heyl3,Budich,Jurcevic,Flaschner,Wang1,Wang2,Cai}. Otherwise, the dynamic quantum phase transition disappears.
So, whether the similarity between the partition function of the classical Ising model and the Loschmidt amplitude for quantum dynamics imply an equivalence between the Ising model with an imaginary magnetic field and non-Hermitian dynamics? If they are equivalent, does this imply that we can directly characterize the Yang-Lee zeros and the Yang-Lee edge singularity of the classical Ising model through quantum dynamic processes?

In this work, we first show that, in the continuum limit, the partition function of the classical 1D Ising model in the presence of an imaginary external magnetic field has an identical mathematical form to the Loschmidt amplitude in non-Hermitian dynamics, with a bounded deviation between the partition function and Loschmidt amplitude when departing from the continuum limit. Furthermore, we demonstrate that, in the general case, the partition function and the Loschmidt amplitude have a completely equivalent form, where the non-Hermitian effective Hamiltonian driving the quantum dynamics takes a more complicated form under the continuum limit. Therefore, a classical Ising model with variable chain length can be mapped to the evolution of quantum dynamics through a classical-quantum correspondence~\cite{Matsumoto,Suzuki,Kogut}. Specifically, we map the classical 1D Ising model to zero-dimensional(0D) non-Hermitian dynamics and the classical two-dimensional(2D) Ising model to the quantum dynamics driven by the Hamiltonian of a quantum 1D Ising model. Additionally, we provide the parameter correspondence between the classical and quantum systems. The results indicate that the Yang-Lee zeros and Yang-Lee edge singularity in the classical system correspond to the dynamic quantum phase transitions and exceptional point~\cite{Berry,Heiss,Miri,Wang3} of the non-Hermitian Hamiltonian in the quantum dynamics, respectively. These conclusions enable the study of Yang-Lee zeros in higher-dimensional Ising models by examining dynamic quantum phase transitions in lower-dimensional systems.

Our work is organized as follows. In Sec.~II, we present the expression for the partition function of the classical 1D Ising model with an imaginary magnetic field, along with the expression for the Yang-Lee zeros of the partition function. In Sec.~III, we rewrite the partition function of the 1D Ising model in the form of the Loschmidt amplitude in 0D non-Hermitian dynamics, demonstrating that the period of the partition function zeros is equivalent to the period of the dynamical quantum phase transition, and we establish the parameter correspondence between the classical and quantum systems. In Sec.~IV, we map classical 2D Ising model to non-Hermitian dynamics driven by the  quantum 1D Ising model. We summarize in Sec.~V.

\section{classical 1D Ising model with imaginary magnetic field}
For the classical 1D Ising model, when a purely imaginary external magnetic field is present~\cite{Fisher2}: $H = -J\sum_{l}  \sigma_l \sigma_{l+1} - ih\sum_{l} \sigma_l$$(J>0, h\in \mathbb{R}, \sigma_{l}=\pm)$, it exhibits Yang-Lee zeros and Yang-Lee edge singularity as shown in Fig.~\ref{fig:fig1}(a). This is most visible by calculating the partition function of the Ising model. The transfer matrix of this model is given by $T = e^{-\beta J}\sigma^x + e^{\beta J}[\cos(\beta h)\mathrm{I}+i\sin(\beta h)\sigma^z]$, and its eigenvalues are $\lambda_{\pm} = e^{\beta J}\cos(\beta h) \pm \sqrt{ -e^{2\beta J}\sin^2(\beta h) + e^{-2\beta J}}$. Under the periodic boundary condition, the partition function is represented as (see Appendix A)
\begin{align}
        Z_{\text{cl}} & = \mathrm{Tr}\exp\left[\beta\sum_l(J\sigma_l\sigma_{l+1}+ih\sigma_l)\right] \notag \\
          & = \mathrm{Tr}[T^N] \notag \\
          & = e^{N\beta J}\left\{\left[\cos(\beta h)+\sqrt{-\sin^2(\beta h)+e^{-4\beta J}}\right]^N \right.  \notag \\
          & \quad +\left. \left[\cos(\beta h)-\sqrt{-\sin^2(\beta h)+e^{-4\beta J}}\right]^N \right\}. \label{eq:partition0}
\end{align}
If $e^{-4\beta J}-\sin^2(\beta h)\leq 0$ is satisfied, the partition function can be rewritten as
\begin{align}
        &Z_{\text{cl}}  \notag \\
          & = 2e^{N\beta J}\left(1-e^{-4\beta J}\right)^{N/2}\cos(N\alpha)  \notag \\
          & = 2e^{N\beta J}\left(1-e^{-4\beta J}\right)^{N/2}\cos\left(N\arccos\frac{\cos(\beta h)}{\sqrt{1-e^{-4\beta J}}}\right)  \notag \\
          & = 2e^{N\beta J}\left(1-e^{-4\beta J}\right)^{N/2}\cos\left(N\arccos\left(\frac{\cos(\beta h)}{\cos(\beta h_c)}\right)\right). \label{eq:partition}
\end{align}
That is, there exists a critical imaginary magnetic field
\begin{align}
h_c = \frac{1}{\beta}\arcsin\left(e^{-2\beta J}\right),   \label{eq:YLcritical}
\end{align}
such that when $|h|>|h_c|$, the partition function exhibits zeros, known as Yang-Lee zeros. The values of the imaginary magnetic field at the zeros are given by
\begin{align}
h = \frac{1}{\beta}\arccos\left\{\sqrt{1-e^{-4\beta J}}\cos\left[\left(m+\frac{1}{2}\right)\frac{\pi}{N}\right]\right\},   \label{eq:YLzeros}
\end{align}
where $m= 0,1,\cdots, N-1$.
So, the Yang-Lee zeros and the critical points are distributed along the imaginary axis of the complex plane formed by the external magnetic field, as shown in Fig.~\ref{fig:fig1}(a). In the thermodynamic limit $N\rightarrow \infty$, the Yang-Lee zeros exhibit a continuous distribution.

% and the anti-$\mathcal{PT}$ symmetric non-Hermitian Hamiltonian is $H = H_{\text{APT}} = ih_x\sigma^x - h_z\sigma^z$. Here, $\sigma^x$, $\sigma^y$, and $\sigma^z$ denote the Pauli matrices, and the anti-$\mathcal{PT}$ symmetric Hamiltonian satisfying $\{\mathcal{PT},H_{\text{APT}}\}=0$ is related to a $\mathcal{PT}$ symmetric Hamiltonian by $H_{\text{APT}}=\pm iH_{PT}$, where $\mathcal{P}=\sigma^x$, $\mathcal{T}$ denotes complex conjugation operation.

%In the following, we will explain in detail that $h_z$ in $H_{\text{APT}}$ is equivalent to the imaginary magnetic field $h$ in the Ising model, $h_x$ corresponds to the critical magnetic field $h_c$, and the exceptional point is equivalent to the Yang-Lee edge critical point. The quantum dynamics governed by $H_{\text{APT}}$ is realized in open quantum systems through postselection, see Appendix B for details.

\section{one-dimensional Ising model with imaginary magnetic field mapped to non-Hermitian dynamics}
In this section, we illustrate that the Yang-Lee zeros and Yang-Lee edge singularity can be implemented in quantum systems on the basis of the classical-quantum correspondence, where a classical system is mapped to a quantum system via the equivalent canonical partition function. Specifically, we consider mapping the classical 1D Ising model with purely imaginary magnetic field to the dynamics driven by a non-Hermitian effective Hamiltonian.

%\subsection{The classical-quantum correspondence in the continuum limit}
We rewrite the partition function of the Ising model in Eq.~(\ref{eq:partition0}) as
\begin{align}
&Z  \nonumber\\
        & = \sum_{\sigma_1 = \pm}\cdots \sum_{\sigma_{N} = \pm}\exp\left[\beta\sum_{k}(J\sigma_{k+1}\sigma_{k}+ ih\sigma_k)\right] \nonumber\\
        &= \left({\frac{1}{2}\sinh(2\theta)}\right)^{-N/2}\sum_{\sigma_1 = \pm }\cdots \sum_{\sigma_{N} = \pm}\prod_{k=1}^{N}\bra{\sigma_{k+1}}e^{\theta\sigma^x}  \nonumber\\
        &  \quad e^{i\beta h\sigma^z}\ket{\sigma_k} \nonumber\\
        & = \left({\frac{1}{2}\sinh(2\theta)}\right)^{-N/2}\sum_{\sigma_1 = \pm }\bra{\sigma_1}\left[e^{\theta\sigma^x}e^{i\beta h\sigma^z}\right]^{N}\ket{\sigma_1}  \nonumber\\
%        & =A^{N}\sum_{\sigma_1 = \pm }\bra{\sigma_1}e^{-it(ih_x\sigma^x-h_z\sigma^z)}\ket{\sigma_1} + \mathrm{Tr}E_{N}
,    \label{eq:Loschmidtclassical}
\end{align}
with $\sigma_{N}=\sigma_1$, and we use the following relation
\begin{align}
        & \sqrt{\cosh(\theta)\sinh(\theta)}\exp\left(\ln\sqrt{\frac{\cosh(\theta)}{\sinh(\theta)}}\sigma_{k+1}\sigma_k + i\beta h\sigma_k \right)  \nonumber\\
        & = \left[\cosh(\theta)\delta_{\sigma_{k+1}\sigma_{k}} + \sinh(\theta)(1-\delta_{\sigma_{k+1}\sigma_{k}})\right]e^{i\beta h\sigma_k}   \nonumber\\
        & = \bra{\sigma_{k+1}}\cosh(\theta)I + \sinh(\theta)\sigma^x\ket{\sigma_k}e^{i\beta h\sigma_k}   \nonumber\\
        & = \bra{\sigma_{k+1}}e^{\theta\sigma^x}e^{i\beta h\sigma^z}\ket{\sigma_k}.
\end{align}
Therefore, Eq.~(\ref{eq:Loschmidtclassical}) can be obtained by simply setting $\ln\sqrt{\frac{\cosh(\theta)}{\sinh(\theta)}}=\beta J$, and $\theta$ can be expressed as $\theta=\frac{1}{2}\ln\frac{1+e^{-2\beta J}}{1-e^{-2\beta J}}$.  Further, Eq.~(\ref{eq:Loschmidtclassical}) can be expressed as 
\begin{align}
Z  &= A^{N}\sum_{\sigma_1 = \pm }\bra{\sigma_1}e^{N\theta\sigma^x+iN\beta h\sigma^z}\ket{\sigma_1}+ \mathrm{Tr}E_{N}  \nonumber\\
       & = A^{N}\sum_{\sigma_1 = \pm }\bra{\sigma_1}e^{-it(ih_x\sigma^x-h_z\sigma^z)}\ket{\sigma_1} + \mathrm{Tr}E_{N} \nonumber\\
        & = A^{N}\mathrm{Tr}\exp(-it H_{\text{APT}}) + \mathrm{Tr}E_{N}, \label{eq:Loschmidtquan}
\end{align}
where the last line above represents the Loschmidt amplitude in a quantum dynamics, with an additional term representing the error, or deviation. The driving Hamiltonian $H_{\text{APT}}=ih_x\sigma^x-h_z\sigma^z$ is an anti-$\mathcal{PT}$ symmetric non-Hermitian Hamiltonian~\cite{Choi,Fang}, satisfying $\{\mathcal{PT},H_{\text{APT}}\}=0$, and is related to a $\mathcal{PT}$ symmetric Hamiltonian by $H_{\text{APT}}=\pm iH_{\text{PT}}$, where $\mathcal{P}=\sigma^x$, $\mathcal{T}$ denotes the complex conjugation. The initial state of the quantum dynamics is a mixed state that is an equal-weight superposition of eigenstates of $\sigma_z$. Then, the coefficeints are given by
    \begin{align}
        A & = \left[\frac{1}{2}\sinh\left(\ln\frac{1+e^{-2\beta J}}{1-e^{-2\beta J}}\right)\right]^{-\frac{1}{2}},\\
              h_x & = \frac{N}{2t}\ln\frac{1+e^{-2\beta J}}{1-e^{-2\beta J}}, \label{eq:parameter11} \\
        h_z & = \frac{N}{t}\beta h.   \label{eq:parameter1}
    \end{align}
The error between the quantum and classical systems is bounded from above as 
\begin{align}
       \left|\mathrm{Tr}E_{N}\right|  &\leq \frac{(\left|N\theta\right| + \left|N\beta h\right|)^3}{3N^2}A^N e^{\left(\left|N\theta\right| + \left|N\beta h\right|\right)}   \nonumber\\
& = N(\left|\theta\right| + \left|\beta h\right|)^3 A^N e^{\left(\left|N\theta\right| + \left|N\beta h\right|\right)},\label{eq:errorterm}
\end{align}
which becomes negligible under the condition $N(\left|\theta\right| + \left|\beta h\right|)^3 \rightarrow 0$~\cite{Matsumoto}. In the following, we introduce the concept of the continuum limit from the perspective of quantum dynamics, and illustrate the condition above is naturally satisfied
in the continuum limit. 

\begin{figure}[tbp]
  \centering
  \includegraphics[width=9cm]{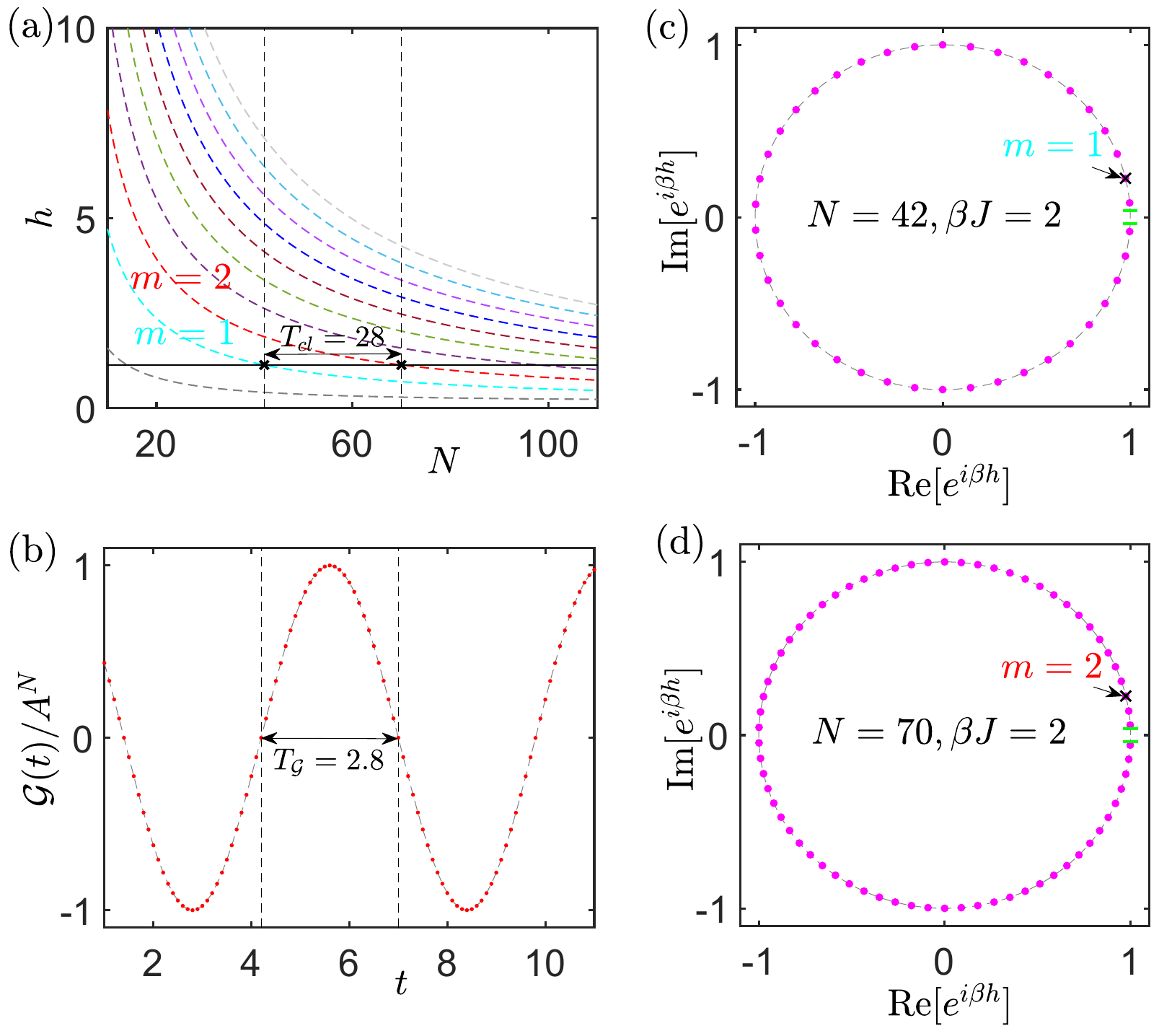}
  \caption{Yang-Lee zeros and edge singularity generated through classical-quantum correspondence. (a) The relation between the imaginary magnetic field and the total number of spins $N$ when the partition function of the Ising model is zero, our numerical results are obtained from Eq.~(\ref{eq:YLzeros}) by taking different values of $m$. The horizontal black line is located at $h=1.137$. (b) The Loschmidt amplitude of the quantum dynamics as a function of $t$, the value of $h_z$ correspond to the magnetic field indicated by the black horizontal line in (a). (c) and (d) The Yang-Lee zeros (magenta dots) lie on the unit circle in the complex plane as a function of $e^{2i\beta h}$. The magnetic field corresponding to the black crosses are the value of the magnetic field indicated by the black horizontal line in (a), while the green segments are the Yang-Lee edge singularities defined in Eq.~(\ref{eq:YLcritical}). For our calculations, we take $\beta=\frac{1}{10}, J=20$. }
  \label{fig:fig2}
\end{figure}

%Here, we discuss the quantum-classical correspondence between the classical 1D ferromagnetic Ising model and a anti-$\mathcal{PT}$ symmetric non-Hermitian system, where this correspondence is based on the equivalence of the canonical partition functions and the Loschmidt amplitude in quantum dynamics. We reconsider the Loschmidt amplitude in Eq. (\ref{eq:Loschmidt}), writing $t$ as $t = t_0 N$ with a variable $N$ and a fixed time interval $t_0$. Based on the minimum measurable time interval in the experiment, each interval $t_0$ is divided into $n_{div}$ sub-intervals, resulting in $N = n_{div}N$ sub-intervals of duration $\frac{t_0}{n_{div}}$. By inserting a complete set between each successive pair of sub-intervals, we obtain a path-integral representation for Loschmidt amplitude
Through Eq.~(\ref{eq:Loschmidtquan}), we establish an equivalence between canonical partition function $Z$ and Loschmidt amplitude in the continuum limit, thereby mapping the classical Ising model to the quantum dynamics. The appearance of zeros in the Loschmidt amplitude indicates the occurrence of dynamic quantum phase transitions during the quantum dynamics. Therefore, the Yang-Lee zeros of the classical Ising model with an imaginary magnetic field can be interpreted as dynamic quantum phase transitions in quantum dynamics driven by a non-Hermitian Hamiltonian, thereby endowing the Yang-Lee zeros with a new physical significance. According to the relation between the parameters in the classical-quantum correspondence of Eqs.~(\ref{eq:parameter11}) and~(\ref{eq:parameter1}), if we fix the parameters $h_x$ and $h_z$ of the Hamiltonian in the quantum dynamics, then simply increasing the length $N$ of the classical Ising chain is equivalent to increasing the evolution time $t$ of the quantum system, ensuring that $\frac{t}{N}$ remains constant. To appreciate the significance of this ratio from the perspective of quantum dynamics, We first express the evolution time $t$ as $t=t_0 n_t$, where $t_0$ is a fixed time interval and $n_t$ is the number of intervals. We then divide each $t_0$ into $n_{div}$ sub-intervals, to ensure that $N=n_{div}n_t$, so that the duration of each sub-interval is equal to $\frac{t_0}{n_{div}}=\frac{t}{N}$ and remains constant. In other words, the evolution time $t$ of the quantum dynamics is divided into $N$ segments, each given by $\frac{t}{N}=\frac{t_0}{n_{div}}$. As $n_{div}\rightarrow\infty$, the length of each sub-interval approaches zero, with $\frac{t}{N}=\frac{t_0}{n_{div}}\rightarrow 0$, which we define as the continuum limit.
Then, according to Eqs.~(\ref{eq:parameter11}) and~(\ref{eq:parameter1}), we have $N(\left|\theta\right| + \left|\beta h\right|)^3 = N\left(\frac{t}{N}\right)^3(\left|h_x\right| + \left|h_z\right|)^3 = \frac{n_{t}t_{0}^3}{n_{div}^2}(\left|h_x\right| + \left|h_z\right|)^3 \rightarrow 0$. Hence the error term in Eq.~(\ref{eq:errorterm}) is negligible in the continuum limit.

Now, we take a different perspective on the relation between the parameters in Eqs.~(\ref{eq:parameter11})-(\ref{eq:parameter1}), where the anti-$\mathcal{PT}$ symmetric Hamiltonian has specific parameters $h_x$ and $h_z$ when $N/t$ takes a fixed value. When $|h_z|>|h_x|$, the Loschmidt amplitude $\mathcal{G}(t)=A^{N}\mathrm{Tr}\exp(-it H_{\text{APT}})=A^{N}\cos(\sqrt{h^{2}_{z}-h^{2}_{x}} t)$ periodically reaches zero $\mathcal{G}(t^{n}_{c})=0$ with a series of critical times $t^{n}_{c}=(n+\frac{1}{2})\pi/\sqrt{h_z^2 - h_x^2}$, corresponding to the dynamic quantum phase transitions, as shown by the green curve in Fig.~\ref{fig:fig1}(b). Thus, the period of the dynamic quantum phase transition is
\begin{equation}
T_\mathcal{G} = \frac{\pi}{\sqrt{h_z^2 - h_x^2}}.  \label{eq:DQPTperiod}
\end{equation}
Conversely, when $|h_z|<|h_x|$, the Loschmidt amplitude has no zeros, and the dynamic quantum phase transition disappears, as shown by the black curve in Fig.~\ref{fig:fig1}(b). Moreover, the right (left) eigenvectors as well as the eigenenergies of the $H_{\text{APT}}$ coalesce at the parameter point $|h_z|=|h_x|$, which is nothing but the exceptional point.

As the size $N$ increases, the partition function $Z$ periodically becomes zero in the form of a cosine function, which is consistent with the periodic variation of the Loschmidt amplitude  $\mathcal{G}(t)$. According to Eq.~(\ref{eq:partition}), the period of the size $N$ at which the zeros of the partition function appear is
\begin{equation}
        T_{\text{cl}}=\frac{\pi}{\arccos\frac{\cos\beta h}{\sqrt{1-e^{-4\beta J}}}}.   \label{eq:period}
\end{equation}
In the continuum limit, $T_{\text{cl}}$ and $T_\mathcal{G}$ are related to each other. %Below, we explain the relation between$T_{\text{cl}}$ and $T_\mathcal{G}$ in more detail.
 Applying the Taylor expansion in terms of $\left(\frac{t}{N}\right)$, we obtain
    \begin{align}
        \frac{1}{T_{\text{cl}}} & = \frac{\arccos\frac{\cos\beta h}{\sqrt{1-e^{-4\beta J}}}}{\pi} \notag \\
    & \simeq \frac{\sqrt{(\frac{t}{N}h_z)^2-(\frac{t}{N}h_x)^2}}{\pi} + \mathcal{O}\left(\frac{t}{N}h_{z}h_{x}\right)^2.
    \end{align}
As such, the error between the period of the dynamic quantum phase transition and the zeros of the partition function of classical Ising model is $\left|\frac{N}{T_{\text{cl}}}-\frac{t}{T_\mathcal{G}}\right|  \simeq \mathcal{O}\left(\frac{t}{N}\right)^2$. We can then rewrite Eq.~(\ref{eq:period}) as $T_{\text{cl}}\simeq\frac{N}{t}T_\mathcal{G}$, as shown in Fig.~\ref{fig:fig2}(a) and (b). For a given imaginary magnetic field, the number of spins for which the partition function vanishes is given by $N= \left(m+\frac{1}{2}\right)\pi/\arccos\frac{\cos\beta h}{\sqrt{1-e^{-4\beta J}}}$. This is illustrated in Fig.~\ref{fig:fig2}(a) by the intersection of the black horizontal line (for $h=1.137$) with the dashed curves. The cyan and red dashed curves represent the results for $m = 1$ and $m = 2$, respectively. Their intersections with the black horizontal line (at $N = 42$ and $N = 70$ respectively) indicate the respective system size $N$ required for the occurrence of Yang-Lee zeros for $h=1.137$. The period $T_{\text{cl}}$ is then calculated according to Eq.~(\ref{eq:period}), yielding $T_{\text{cl}} = 28$.
%The general relation between the imaginary magnetic field $h$ and total spin number $N$ for $Z=0$ is Eq.~(\ref{eq:YLzeros}), as shown by colored dashed curves in Fig.~\ref{fig:fig2}(a).

In Fig.~\ref{fig:fig2}(b), the parameters $h_x$ and $h_z$ used for the Loschmidt amplitude are mapped from the corresponding classical Ising model parameters of the black horizontal line in Fig.~\ref{fig:fig2}(a) using Eqs.~(\ref{eq:parameter11}) and~(\ref{eq:parameter1}).
The magenta dots on the unit circles in Fig.~\ref{fig:fig2}(c)(d) represent the Yang-Lee zeros for $N=42$ and $N=70$, respectively. Their corresponding magnetic-field values $h$ are given by the intersections of the vertical dashed lines (located at $N=42$ and $N=70$ respectively) with the dashed curves.
The short green segments in Fig.~\ref{fig:fig2}(c)(d) indicate $h_c$ from Eq.~(3), which are
the critical magnetic fields corresponding to the Yang-Lee edge singularities.
The black crosses in Fig.~\ref{fig:fig2}(c)(d) represent the points where $h=1.137$, which are consistent with the black crosses in Fig.~\ref{fig:fig2}(a).
It can be observed that the black cross in Fig.~\ref{fig:fig2}(c) coincides with the second root (with $m=1$), and the one in Fig.~\ref{fig:fig2}(d) coincides with the third root (with $m=2$).
%It can be observed that for different values of $N$, the same magnetic field value corresponds to different values of $m$.

In the classical Ising model, the Yang-Lee edge singularity is accompanied by anomalous scaling laws with no counterparts in unitary critical phenomena~\cite{Fisher2,Matsumoto}. The magnetization density can be expressed as $m=-\partial_{h}f$, which scales in the vicinity of the critical point as $m\propto\Delta h^{-\frac{1}{2}}$, where $\Delta h=h-h_c$, and the free energy density $f=-\lim_{N\rightarrow\infty}\frac{1}{\beta N}\ln Z$. By differentiating the magnetization density with respect to the magnetic field, one can obtain the scaling law for the magnetic susceptibility $\chi=\frac{d m}{d h}\propto\Delta h^{-\frac{3}{2}}$.
Following the mapping scheme above, the anomalous scaling behavior also has its counterparts in the mapped quantum dynamics.
Setting $\beta=\frac{t}{N}$, we redefine the parameters in Eqs.~(\ref{eq:parameter11}) and~(\ref{eq:parameter1}) as
\begin{align}
h_{z}=h, \quad h_x = \frac{1}{2\beta}\ln\frac{1+\sin(\beta h_c)}{1-\sin(\beta h_c)} \simeq h_c, \label{eq:parameter2}
\end{align}
where $h_c$ is the critical magnetic field and the error between $h_c$ and $h_x$ is $|h_x - h_c|  \simeq \mathcal{O}\left(\frac{t}{N}\right)^2$. Therefore, the critical magnetic field $h=h_c$ in the classical system corresponds to the exceptional point $h_z=h_x$ of the non-Hermitian Hamiltonian in the quantum system. A similar scaling behavior exists as $m'=\partial_{h_z}g\propto\tau^{-\frac{1}{2}}$ and $\chi'=\frac{d m'}{d h}_{z}\propto\tau^{-\frac{3}{2}}$ in the vicinity of the exceptional point, where $\tau = h_z - h_x$. The free energy density of the quantum system is given by $g=-\lim_{t\rightarrow\infty}\frac{1}{t}\ln \mathcal{G}$. The period $T_{\text{cl}} $ in Eq.~(\ref{eq:period}) scales in the vicinity of the exceptional point as~\cite{Fisher2,Fisher3,Xiao}
\begin{equation}
        T_{\text{cl}} \sim \tau^{-\frac{1}{2}}.
\end{equation}
Hence, the anomalous scaling laws in the Yang-Lee edge singularity, which are challenging to observe due to the requirement of an imaginary magnetic field, can be revealed instead by measuring the scaling behavior near the exceptional point in non-Hermitian quantum dynamics. Moreover, in the vicinity of the exceptional point, non-Hermitian quantum dynamics exhibit two additional scaling laws that have not been discussed in the context of the classical Ising model. The critical phenomena of these scaling behaviors are discussed in detail in Refs.~\cite{Matsumoto,Gao}.

So far, we explained how the classical Ising model with an imaginary magnetic field and variable system size can be mapped to non-Hermitian dynamics using the formal similarity between the partition function and the Loschmidt amplitude.
As such, the appearance of Yang-Lee zeros indicates the occurrence of dynamic quantum phase transitions in the corresponding quantum system.
Specifically, for a given imaginary magnetic field, the system sizes $N$ for the occurrence of  Yang-Lee zeros is equivalent to the critical times $t^{n}_{c}$ at which dynamic quantum phase transitions occur in the quantum dynamics with a fixed parameter $h_x$ and $h_z$.
Conversely, if $N$ is fixed, there exist $N$ values of the imaginary magnetic field that give rise to Yang-Lee zeros.
When mapped to quantum dynamics, this corresponds to a fixed time $t$, where different values of $h_z$ (with $h_x$ fixed) lead to the occurrence of dynamic quantum phase transitions.
To further understand the physical meaning behind this classical-quantum mapping, we recall Ref.~\cite{Matsumoto}, where the canonical partition function of a $\mathcal{PT}$-symmetric non-Hermitian quantum system (with a $2\times2$ Hamiltonian) is mapped to the partition function of a classical 1D Ising model. Consequently, a dynamic quantum phase transition can also be interpreted as the zero of the partition function of a finite-temperature quantum system, where the dynamic evolution time $t$ can be mapped to $\beta$ (the inverse temperature $T$).

Finally, we consider a more general case, where by setting aside the constraint of the continuum limit, the partition function in Eq.~(\ref{eq:Loschmidtclassical}) can be explicitly written as (see Appendix B)
\begin{align}
Z &= A^{N}\sum_{\sigma_1 = \pm }\bra{\sigma_1}\left[e^{\theta\sigma^x}e^{i\beta h\sigma^z}\right]^{N}\ket{\sigma_1}  \nonumber\\
& = A^{N}\sum_{\sigma_1 = \pm }\bra{\sigma_1}\exp(-it H_{\text{non}})\ket{\sigma_1}  \nonumber\\
& = \mathcal{G^{'}}(t),
\label{eq:partition2}
\end{align}
with the non-Hermitian Hamiltonian $H_{\text{non}}$ given by a $2\times 2$ matrix.
Thus, there exists a strict mathematical equivalence between the canonical partition function $Z$ and the Loschmidt amplitude $\mathcal{G^{'}}(t)$. This means that the partition function of the Ising model can always be mapped to a Loschmidt amplitude in quantum dynamics driven by some non-Hermitian Hamiltonian. The inverse mapping process however, is not always possible without resorting to the continuum limit (see Appendix B).

\begin{figure*}[tbp]
  \centering
  \includegraphics[width=13.5cm]{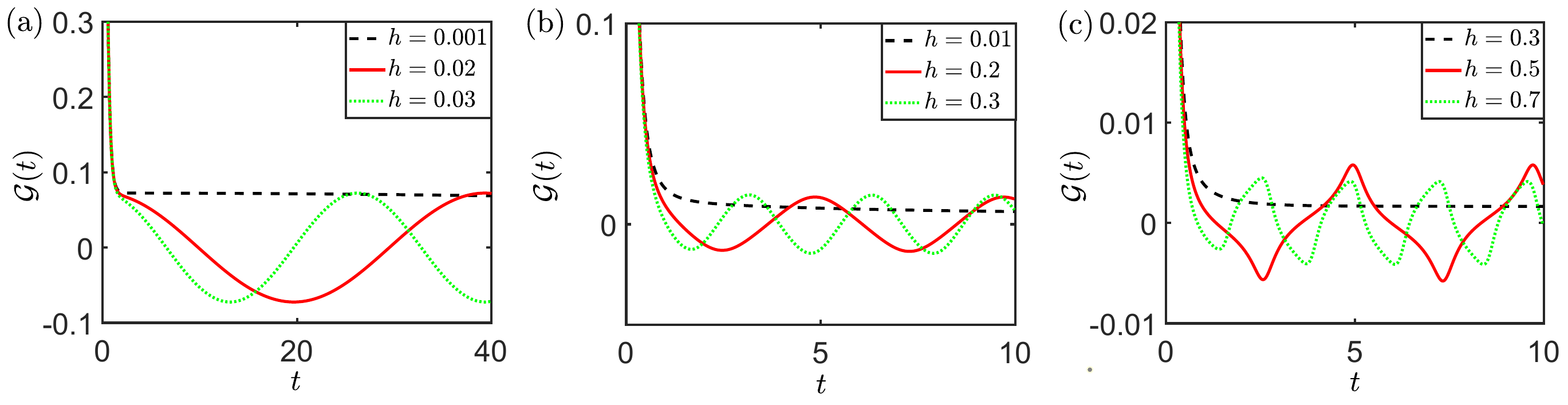}
  \caption{The Loschmits amplitude for different value of $g$: (a) $g=0.1<1$;(b) $g=1$;(c) $g=2>1$}
  \label{fig:fig3}
\end{figure*}

\section{two-dimensional Ising model mapped to 1D non-Hermitian dynamics}
In this section, we study how a classical 2D Ising lattice  is mapped to non-Hermitian dynamics driven by the Hamiltonian of a quantum 1D Ising model. The Hamiltonian for the 2D Ising model can be written as~\cite{Onsager,CNYang}
\begin{align}
        H_{2D} = \sum_{n,m=1,1}^{N,M}(-J_1\sigma_{n,m}\sigma_{n+1,m}-J_2\sigma_{n,m}\sigma_{n,m+1} - ih_{cl}\sigma_{n,m}). \label{eq:Hamiltonian2D}
\end{align}
Similar to the classical-quantum correspondence in the previous section, we use path-integral representation to express the partition function (see Appendix C)
\begin{align}
        Z &  =\mathrm{Tr}e^{-\beta_{cl} H_{2D}}    \nonumber\\
&  \sim \bra{\psi_0}\exp (-itH_1)\ket{\psi_0} = \frac{1}{2^N}\mathrm{Tr}\exp(-it H_1),  \label{eq:partition2D}
\end{align}
where the Hamiltonian of the quantum Ising model with both transverse and longitudinal fields has the form~\cite{Elliott,Pfeuty,Stinchcombe,Heyl3}
\begin{equation}
H_1 = iJ\left(\sum_{n=1}^{N}\sigma^z_n\sigma^z_{n+1}+g\sum_{n=1}^{N}\sigma^x_n\right) - h\sum_{n=1}^{N}\sigma^z_n.
\label{eq:Hamiltonian1}
\end{equation}
Here the initial state $\ket{\psi_0}$ of the quantum dynamics is a mixed state that is an equal-weight superposition of all eigenstates of the Hamiltonian $H_1$. Then, the coefficients are given by
\begin{align}
        J  & = \frac{M}{t}\beta_{cl}J_1    \notag \\
        Jg & = \frac{M}{2t}\ln\frac{1+e^{-2\beta_{cl} J_2}}{1-e^{-2\beta_{cl} J_2}}  \notag \\
        h & = \frac{M}{t}\beta_{cl}h_{cl},   \label{eq:parameter2}
\end{align}
where the ratio $\frac{t}{M}=\frac{t_0}{n_{div}}\rightarrow 0$ means the continuum limit.
Thus, by utilizing the equivalence between the canonical partition function of a 2D Ising model and the Loschmidt amplitude in the quantum dynamics of a 1D Ising model, we map a 2D Ising lattice with variable system size to the quantum dynamics of a 1D Ising chain, where the continuous increase in the longitudinal size  corresponds to the increase in evolution time. The zeros of the partition function in the 2D Ising lattice correspond to the dynamic quantum phase transitions in the 1D Ising chain.

\begin{figure}[tbp]
  \centering
  \includegraphics[width=8.5cm]{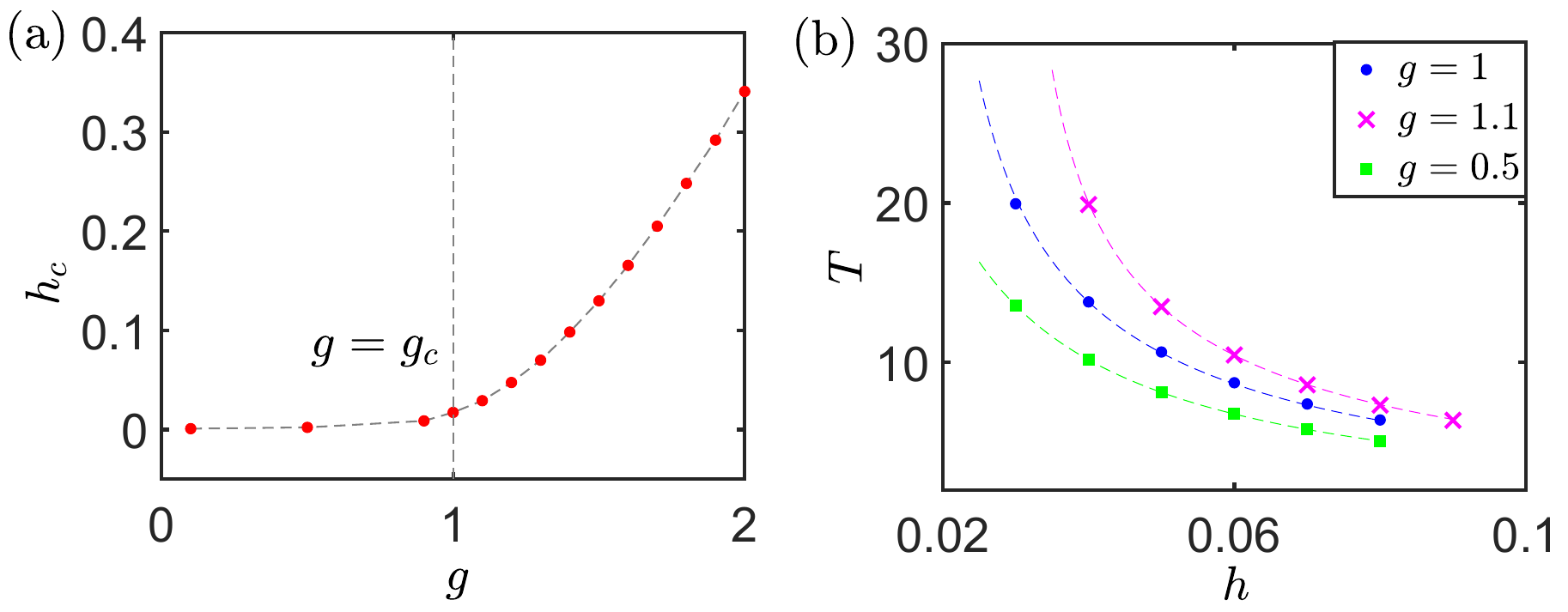}
  \caption{(a) The critical imaginary magnetic field as a function of $g$, there is a sudden growth in $g=1$;     (b) The period of DQPT as a function of h, we fit the result and obtain the critical phenomena of T.}
  \label{fig:fig4}
\end{figure}

Since the partition function of the 2D Ising model with external magnetic field in Eq.~(\ref{eq:partition2D}) lacks an analytical solution, we study the Yang-Lee zeros and the Yang-Lee edge singularity of the 2D Ising model with $N=8$ by numerically fitting the quantum dynamics driven by the Hamiltonian $H_1$. The variation of the Loschmidt amplitudes over time under different parameters $g$ and magnetic fields $h$ are shown in Fig.~\ref{fig:fig3}. As we can see, the dynamic quantum phase transition happens with a period just like the 0D case.  Under the same parameter $g$, the stronger the imaginary magnetic field, the shorter the period, as shown in Fig.~\ref{fig:fig3} and Fig.~\ref{fig:fig4}(b).  According to numerical fitting, the period of the dynamic quantum phase transition satisfies
\begin{equation}
T = \frac{\alpha}{\sqrt{h^2-h_c^2(J,g,N)}},
\end{equation}
with $\alpha$ is a fitting parameter. The period $T$ is similar to Eq.~(\ref{eq:DQPTperiod}), and the scaling law $T\sim\tau^{-\frac{1}{2}}$ is the same as the previous section. There is also a critical magnetic field $h_c$ such that only under the condition $|h|>h_c$ that the dynamic quantum phase transition happens. The results in Fig.~\ref{fig:fig3} and Fig.~\ref{fig:fig4}(a) also indicate that different parameters $g$ have different critical magnetic fields $h_c$. Specifically, when $g>1$, there exists a critical imaginary magnetic field of a finite value, while when $g<1$, the critical imaginary magnetic field becomes very small. We know that the phase transition point of the 1D transverse-field Ising model is $|g|=1$. As the parameter changes from $|g|>1$ to $|g|<1$, the system undergoes a transition from the disordered phase to the ordered phase. As shown in Fig.~\ref{fig:fig4}(a), this phase transition point is also the critical magnetic field transition point. As the value of $g$ decreases (in a finite-size system), the critical magnetic field continuously approaches zero.

To further understand these results, we map the quantum dynamics driven by the Hamiltonian $H_1$ back to the 2D Ising model. According to the parameter mapping relation in the classical-quantum correspondence in Eq.~(\ref{eq:parameter2}), the phase transition point $|g|=1$ of the 1D transverse field Ising model corresponds to $J_{2,c}=-\frac{1}{2\beta_{cl}}\ln\tanh(\frac{t}{M}J)$ in the 2D Ising model. During the transition from $g>1$ to $g<1$, the parameters of the 2D system change from $J_2>J_{2,c}$ to $J_2<J_{2,c}$, while the system transitions from the disordered phase to the ordered phase. According to the Yang-Lee theory, when the system is in the ordered phase, the minimum imaginary external field corresponding to the zeros of the partition function in the thermodynamic limit also becomes very small. When the 2D system is in the disordered phase, there exists a critical value for the imaginary external magnetic field $h_{cl,c}=\frac{it}{M}h_c$, and the zeros of the partition function only exist in $|h_{cl}|>|h_{cl,c}|$.

\section{Summary}
To summarize, we show that the Yang-Lee zeros in the classical Ising models can be mapped to the dynamic quantum phase transitions in non-Hermitian quantum systems, and the Yang-Lee edge singularity can be mapped to the exceptional point of the non-Hermitian effective Hamiltonian. We confirm this understanding by demonstrating that the partition function of the classical Ising model and the Loschmidt amplitude in the mapped quantum dynamics have a completely equivalent mathematical form. This equivalence allows for the experimentally challenging problem of imaginary magnetic fields to be transformed into a simpler problem of nonunitary evolution in open quantum systems.
Moreover, the classical-to-quantum mapping reduces the dimensionality of the Ising model under investigation, enabling us to study the Yang-Lee zeros and phase transition problems of higher-dimensional classical Ising models through the quantum dynamics of low-dimensional systems.

In a recent experimental study~\cite{Gao}, the Yang-Lee criticality has been probed in the 0D non-Hermitian dynamics using single photons. We expect that the setup therein can be extended to illustrate our results.
Further, the 1D non-Hermitian Ising model discussed in Sec.~IV can be implemented using Rydberg atoms following Ref.~\cite{Shen1}. Specifically, the transverse field can be generated by a microwave field, and the imaginary term can be introduced through laser-induced atom loss~\cite{Lapp1,Ren1}.
 
%To verify our scheme driven by the 1D quantum Ising model, we suggest to follow the proposal in reference ~\cite{Shen1}, where the authors  use Rydberg-dressing process to produce the effective interaction between spins, a microwave field to generate the transverse field, a laser to introduce the atom loss~\cite{Lapp1,Ren1} as imaginary fields and Floquet driving to simulate the dynamics. In this way, the experimental implementation difficulty can be reduced by lowering the dimensionality of the simulated quantum system.

\appendix

\section{detailed derivation of the partition function for 1D Ising model}
We rewrite the Hamiltonian in Eq.~(\ref{eq:partition0}) as a sum over bonds
    \begin{equation}
        H = -\sum_{l} E(\sigma_l,\sigma_{l+1}),
    \end{equation}
    where $E(\sigma_l,\sigma_{l+1}) = -J\sigma_{l}\sigma_{l+1} - \frac{h}{2}(\sigma_{l}+\sigma_{l+1})$ is the energy of the bond between sites $l$ and $l+1$.(Note that $E(\sigma_l, \sigma_{l+1})$ takes on four possible values, since there are four spin configurations on sites $l$ and $l+1$: $++$, $+-$, $-+$,and $--$.) Expressing the Hamiltonian in this form allows the Boltzmann weight, $e^{-\beta H}$, to be factorized into a product
\begin{align}
          & e^{-\beta H(\sigma_1,\sigma_2,...,\sigma_N)}  \notag \\
          & = e^{-\beta E(\sigma_1,\sigma_2) - \beta E(\sigma_2,\sigma_3) -...- \beta E(\sigma_N,\sigma_1)} \notag \\
          & = e^{-\beta E(\sigma_1,\sigma_2)}e^{-\beta E(\sigma_2,\sigma_3)...e^{-\beta E(\sigma_N,\sigma_1)}}
  \notag \\
          & = T_{\sigma_1,\sigma_2}T_{\sigma_2,\sigma_3}...T_{\sigma_{N-1},\sigma_{N}}T_{\sigma_{N},\sigma_1};
\end{align}
that is, there's just a factor of $T$ for each of the bonds in the Hamiltonian. For a bond between a spin $\sigma$ and another spin $\sigma'$, each spin can take one of two values ($+1$ or $-1$). This results in a $2\times2$ transfer matrix $T$, containing four entries. Explicitly, the transfer matrix $T$ can be written as
 \begin{align}
        T &= \begin{pmatrix}
            e^{-\beta E(+1,+1)} & e^{-\beta E(+1,-1)} \\
            e^{-\beta E(-1,+1)} & e^{-\beta E(-1,-1)}
        \end{pmatrix}  \notag \\
          &= \begin{pmatrix}
            e^{\beta J + \beta ih} & e^{-\beta J} \\
            e^{-\beta J} & e^{\beta J-\beta ih}
        \end{pmatrix}.
 \end{align}

With a concise expression for $e^{-\beta H}$ now in hand, we return to the task of determining the partition function $Z$. This requires summing over all possible configurations of the system, as outlined in the following calculation
\begin{align}
          & \mathrm{Tr} e^{-\beta H(\sigma_1,\sigma_2,...,\sigma_N)}  \notag \\
          & = \sum_{\sigma_1 = \pm}\sum_{\sigma_2 = \pm}\cdots \sum_{\sigma_{N} = \pm} T_{\sigma_1,\sigma_2}T_{\sigma_2,\sigma_3}...T_{\sigma_{N-1},\sigma_{N}}T_{\sigma_{N},\sigma_1}
  \notag \\
          & = \sum_{\sigma_1 = \pm}\sum_{\sigma_2 = \pm}\cdots \sum_{\sigma_{N} = \pm} \bra{\sigma_1}T\ket{\sigma_2}\bra{\sigma_2}T\ket{\sigma_3}...\bra{\sigma_{N}}T\ket{\sigma_1}  \notag \\
          & = \sum_{\sigma_1 = \pm}\bra{\sigma_1}T^N\ket{\sigma_1}   \notag \\
          & = \mathrm{Tr}[T^N] .
\end{align}

\begin{widetext}
\section{classical-quantum correspondence in the general cases}
The non-Hermitian Hamiltonian in Eq.~(\ref{eq:partition2}) of the main text has the form (using the Baker-Campbell-Hausdorff formula)
%\begin{align}
%H_{\text{non}}& =  \frac{i}{t}\ln\left[e^{\theta\sigma^x}e^{i\beta h\sigma^z}\right]^{N}   \nonumber\\
%& = \frac{iN}{t}\ln\left[\cosh(\theta)\cos(\beta h)\mathrm{I} + \sinh(\theta)\cos(\beta h)\sigma^x - \sinh(\theta)\sin(\beta h)\sigma^y + i\cosh(\theta)\sin(\beta h)\sigma^z\right] \nonumber\\
%& = \frac{iN}{t}\left\{\frac{\tan^{-1}(\lambda/(\cosh(\theta)\cos(\beta h)))}{\lambda}[\sinh(\theta)\cos(\beta h)\sigma^x - \sinh(\theta)\sin(\beta h)\sigma^y + i\cosh(\theta)\sin(\beta h)\sigma^z] \right.  \nonumber\\
%&  \quad +\left.\ln(\sqrt{(\cosh(\theta)\cos(\beta h))^2+(\sinh(\theta))^2})\mathrm{I} \right\}   \nonumber\\
%& =h^{'}_{x}(J,h)\sigma^{x}+h^{'}_{y}(J,h)\sigma^{y}+h^{'}_{z}(J,h)\sigma^{z}+\ln(\sqrt{(\cosh(\theta)\cos(\beta h))^2+(\sinh(\theta))^2})\mathrm{I}. \label{eq:FloquetHTI}
%
%\end{align}

\begin{align}
H_{\text{non}}& =  \frac{i}{t}\ln\left[e^{\theta\sigma^x}e^{i\beta h\sigma^z}\right]^{N}   \nonumber\\
& = \frac{iN}{t}(\theta\sigma^x + i\beta h\sigma^z +\frac{1}{2}[\theta\sigma^x , i\beta h\sigma^z]    + \frac{1}{12}[\theta\sigma^x , [\theta\sigma^x , i\beta h\sigma^z]]  - \frac{1}{12}[i\beta h\sigma^z , [\theta\sigma^x , i\beta h\sigma^z]]+\cdots) \nonumber\\
& =h^{'}_{x}(J,h)\sigma^{x}+h^{'}_{y}(J,h)\sigma^{y}+h^{'}_{z}(J,h)\sigma^{z}. \label{eq:FloquetHTI}
\end{align}

Here the invertible matrix $e^{\theta\sigma^x}e^{i\beta h\sigma^z}$ ensures the feasibility of logarithmic operations, and the coefficients take on a relatively complex form
%    \begin{align}
%        \lambda(\theta,h) & = \sqrt{(\sinh(\theta)\cos(\beta h))^2+(\sinh(\theta)\sin(\beta h))^2-(\cosh(\theta)\sin(\beta h))^2} \notag \\  \label{eq:parameter30} \\
%        h^{'}_{x}(\theta,h) & = \frac{iN}{t}\left[\frac{\tan^{-1}(\lambda/(\cosh(\theta)\cos(\beta h)))}{\lambda}\sinh(\theta)\cos(\beta h)\right] \notag \\  \label{eq:parameter31} \\
%        h^{'}_{y}(\theta,h) & = -\frac{iN}{t}\left[\frac{\tan^{-1}(\lambda/(\cosh(\theta)\cos(\beta h)))}{\lambda}\sinh(\theta)\sin(\beta h)\right] \notag \\  \label{eq:parameter32} \\
%        h^{'}_{z}(\theta,h) & = -\frac{N}{t}\left[\frac{\tan^{-1}(\lambda/(\cosh(\theta)\cos(\beta h)))}{\lambda}\cosh(\theta)\sin(\beta h)\right].      \label{eq:parameter3}
%    \end{align}

    \begin{align}
        h^{'}_{x}(J,h) & = \frac{iN}{t}\left[\theta + \frac{1}{3}\theta(\beta h)^2 + \cdots\right] \notag \\
& = \frac{iN}{t}\left[\frac{1}{2}\ln\frac{1+e^{-2\beta J}}{1-e^{-2\beta J}}  + \frac{1}{6}(\beta h)^2\ln\frac{1+e^{-2\beta J}}{1-e^{-2\beta J}} + \cdots\right] \label{eq:parameter31} \\
        h^{'}_{y}(J,h) & = \frac{iN}{t}\left[\theta\beta h + \frac{2}{45}\theta^3(\beta h)^3 + \cdots\right] \notag \\
& = \frac{iN}{2t}\left[\beta h\ln\frac{1+e^{-2\beta J}}{1-e^{-2\beta J}}  +  \frac{1}{180}(\beta h)^3\left(\ln\frac{1+e^{-2\beta J}}{1-e^{-2\beta J}}\right)^3\cdots\right]  \label{eq:parameter32} \\
        h^{'}_{z}(J,h) & = -\frac{N}{t}\left[\beta h + \frac{1}{3}\theta^2\beta h + \cdots\right]   \notag \\
& = -\frac{N}{t}\left[\beta h + \frac{1}{12}\left(\ln\frac{1+e^{-2\beta J}}{1-e^{-2\beta J}}\right)^2\beta h + \cdots\right].   \label{eq:parameter3}
    \end{align}

 According to the parameter relation in Eqs.~(\ref{eq:parameter31}) to (\ref{eq:parameter3}), increasing the length $N$ of the classical Ising chain is also equivalent to increasing the evolution time $t$ of the quantum system. Although the non-Hermitian Hamiltonian $H_{\text{non}}$ driving the system may take a relatively complicated form, such a $H_{\text{non}}$ always exists in principle. If the coefficients of the Hamiltonian satisfy $(h^{'}_{x})^2+(h^{'}_{y})^2+(h^{'}_{z})^2>0$, then $\sqrt{(h^{'}_{x})^2+(h^{'}_{y})^2+(h^{'}_{z})^2}=\frac{N}{t}\arccos\left(\frac{\cos(\beta h)}{\cos(\beta h_c)}\right)$. In addition, the Loschmidt amplitude $\mathcal{G^{'}}(t)=A^{N}\mathrm{Tr}\exp(-it H_{\text{non}})=A^{N}\cos(\sqrt{(h^{'}_{x})^2+(h^{'}_{y})^2+(h^{'}_{z})^2}t)$ periodically reaches zero $\mathcal{G^{'}}(t^{n}_{c})=0$ with a series of critical times $t^{n}_{c}=(n+\frac{1}{2})\pi/\sqrt{(h^{'}_{x})^2+(h^{'}_{y})^2+(h^{'}_{z})^2}$, corresponding to dynamic quantum phase transitions. Thus, the period of the dynamic quantum phase transition in this case is
\begin{equation}
T_\mathcal{G^{'}} = \frac{\pi}{\sqrt{(h^{'}_{x})^2+(h^{'}_{y})^2+(h^{'}_{z})^2}},  \label{eq:DQPTperiod1}
\end{equation}
which is equivalent to $T_{\text{cl}}$ as $\frac{N}{T_{\text{cl}}}=\frac{t}{T_\mathcal{G^{'}}}$. Conversely, when $(h^{'}_{x})^2+(h^{'}_{y})^2+(h^{'}_{z})^2<0$, the Loschmidt amplitude has no zeros, and the parameter point $(h^{'}_{x})^2+(h^{'}_{y})^2+(h^{'}_{z})^2=0$ is an exceptional point of $H_{\text{non}}$.

In the continuum limit, as we mentioned in the main text, we have $N/t\to \infty$, for given parameters $h_x',h_y'$ and $h_z'$. According to Eqs.~(\ref{eq:parameter31}) to (\ref{eq:parameter3}), we have $\theta,h\to0$ and thus recover the results in the continuum limit Eqs.~(\ref{eq:parameter11}) and (\ref{eq:parameter1}).

\end{widetext}

\begin{widetext}
\section{detailed derivation of mapping the partition function of 2D Ising model to the Loschmidt amplitude}

Here, we present the detailed derivation of mapping the partition function of the classical 2D Ising model to the Loschmidt amplitude in the quantum dynamics of a 1D Ising model.
We rewrite the partition function of the classical 2D Ising model in Eq.~(\ref{eq:partition2D}) as
\begin{align}
Z& =  \mathrm{Tr}\exp[\sum_{m=1}^{M}\sum_{n=1}^{N}(\beta_{cl} J_1\sigma_{n,m}\sigma_{n+1,m} + \beta_{cl} J_2\sigma_{n,m}\sigma_{n,m+1} + i\beta_{cl} h_{cl}\sigma_{n,m})]   \nonumber\\
& = B^N \sum_{\sigma_1 = \pm }\cdots \sum_{\sigma_{M} = \pm}\prod_{m=1}^{M}\bra{\sigma_{m+1}}\exp\left(\sum_{n=1}^{N}\theta^{'}\sigma_{n}^{x}\right)\exp\left(\sum_{n=1}^{N}\beta_{cl} J_{1}\sigma_{n}^{z}\sigma_{n+1}^{z}\right)\exp\left(i\sum_{n=1}^{N}\beta_{cl} h_{cl}\sigma_{n}^{z}\right)\ket{\sigma_m}  \nonumber\\
& = B^N \mathrm{Tr}\left[\exp\left(\sum_{n=1}^{N}\theta^{'}\sigma_{n}^{x}\right)\exp\left(\sum_{n=1}^{N}\beta_{cl} J_{1}\sigma_{n}^{z}\sigma_{n+1}^{z}\right)\exp\left(i\sum_{n=1}^{N}\beta_{cl} h_{cl}\sigma_{n}^{z}\right)\right]^N, \label{eq:partition2D_1}
\end{align}
with $B = \left[\frac{1}{2}\sinh\left(\ln\frac{1+e^{-2\beta_{cl} J_2}}{1-e^{-2\beta_{cl} J_2}}\right)\right]^{-\frac{1}{2}}$. Here we use the following evaluation  of a matrix element
\begin{align}
        & \sqrt{\cosh(\theta^{'})\sinh(\theta^{'})}\exp\left(\sum_{n=1}^{N}\beta_{cl} J_1\sigma_{n,m}\sigma_{n+1,m}+\sum_{n=1}^{N}\ln\sqrt{\frac{\cosh(\theta^{'})}{\sinh(\theta^{'})}}\sigma_{n,m}\sigma_{n,m+1} + i\sum_{n=1}^{N}\beta_{cl} h_{cl}\sigma_{n,m} \right)  \nonumber\\
        & = \bra{\sigma_{m+1}}\exp\left(\sum_{n=1}^{N}\theta^{'}\sigma_{n}^{x}\right)\ket{\sigma_m}\exp\left(\sum_{n=1}^{N}\beta_{cl} J_{1}\sigma_{n,m}\sigma_{n+1,m}+i\sum_{n=1}^{N}\beta_{cl} h_{cl}\sigma_{n,m}\right)   \nonumber\\
        & = \bra{\sigma_{m+1}}\exp\left(\sum_{n=1}^{N}\theta^{'}\sigma_{n}^{x}\right)\exp\left(\sum_{n=1}^{N}\beta_{cl} J_{1}\sigma_{n}^{z}\sigma_{n+1}^{z}\right)\exp\left(i\sum_{n=1}^{N}\beta_{cl} h_{cl}\sigma_{n}^{z}\right)\ket{\sigma_m}.
\end{align}
Therefore, Eq.~(\ref{eq:partition2D_1}) can be obtained by simply setting $\ln\sqrt{\frac{\cosh(\theta^{'})}{\sinh(\theta^{'})}}=\beta_{cl}J_2$, and $\theta^{'}$ can be expressed as $\theta^{'}=\frac{1}{2}\ln\frac{1+e^{-2\beta_{cl} J_2}}{1-e^{-2\beta_{cl} J_2}}$. Further, Eq.~(\ref{eq:partition2D_1}) can be expressed as
\begin{align}
        Z &  =B^N \mathrm{Tr}\left[\exp\left(\sum_{n=1}^{N}M\theta^{'}\sigma_{n}^{x}+\sum_{n=1}^{N}M\beta_{cl} J_{1}\sigma_{n}^{z}\sigma_{n+1}^{z}+i\sum_{n=1}^{N}M\beta_{cl} h_{cl}\sigma_{n}^{z} \right)\right] + \mathrm{Tr}E_N  \nonumber\\
&  \sim \bra{\psi_0}\exp (-itH_1)\ket{\psi_0} = \frac{1}{2^N}\mathrm{Tr}\exp(-it H_1),
\end{align}
By comparing with the Hamiltonian  in Eq.~(\ref{eq:Hamiltonian1}), the relation between the coefficients can be obtained as given in Eq.~(\ref{eq:parameter2}). Here, the error between the classical and quantum systems is
\begin{align}
       \left|\mathrm{Tr}E_{N}\right|  \leq  M(|\theta^{'}| + \left|\beta_{cl} J_1\right| + \left|\beta_{cl} h_{cl}\right|)^3 B^N \exp{\left(|M\theta^{'}| + \left|M\beta_{cl} J_1\right| + \left|M\beta_{cl} h_{cl}\right|\right)},
\end{align}
When the continuous limit condition is satisfied, this error term is also negligible compared to the Loschmidt amplitude: $n_{div}\rightarrow\infty$, $M(|\theta^{'}| + \left|\beta_{cl} J_1\right| + \left|\beta_{cl} h_{cl}\right|)^3=M\left(\frac{t}{M}\right)^3(|J|+|Jg|+|h|)^3=\frac{n_t t_{0}^3}{n_{div}^2}(|J|+|Jg|+|h|)^3\rightarrow 0$.
\end{widetext}

%***********************************
\begin{acknowledgments}
This work is supported by the National Natural Science Foundation of China (Grant No. 12374479), and by the Innovation Program for Quantum Science and Technology (Grant No. 2021ZD0301205).
\end{acknowledgments}

\end{document}